\documentclass[aip,amsmath,amssymb,reprint]{revtex4-1}
\usepackage{mathptmx}
\usepackage[T1]{fontenc}
\usepackage[utf8]{inputenc}
\usepackage{color}
\usepackage{bm}
\usepackage{amsmath}
\usepackage{graphicx}
\usepackage[unicode=true,pdfusetitle,
 bookmarks=true,bookmarksnumbered=false,bookmarksopen=false,
 breaklinks=false,pdfborder={0 0 1},backref=false,colorlinks=false]
 {hyperref}
\hypersetup{
 colorlinks=true,allcolors=blue}

\makeatletter

\providecommand{\tabularnewline}{\\}

\usepackage{dcolumn}
\usepackage{bm}

\usepackage[utf8]{inputenc}
\usepackage[T1]{fontenc}
\usepackage{mathptmx}
\usepackage{mathtools}

\makeatother

\begin{document}
\title{Nonequilibrium effects of cavity leakage and vibrational dissipation in thermally-activated polariton chemistry} 

\author{Matthew Du}  

\author{Jorge A. Campos-Gonzalez-Angulo}  

\author{Joel Yuen-Zhou} 
\email{joelyuen@ucsd.edu} 
\homepage{http://yuenzhougroup.ucsd.edu} 
\affiliation{Department of Chemistry and Biochemistry, University of California San Diego, La Jolla, California 92093, USA} 

\date{\today}
\begin{abstract}
In vibrational strong coupling (VSC), molecular vibrations strongly
interact with the modes of an optical cavity to form hybrid light-matter
states known as vibrational polaritons. Experiments show that the
kinetics of thermally activated chemical reactions can be modified
by VSC. Transition-state theory, which assumes that internal thermalization
is fast compared to reactive transitions, has been unable to explain
the observed findings. Here, we carry out kinetic simulations to understand
how dissipative processes, namely those that VSC introduces to the
chemical system, affect reactions where internal thermalization and
reactive transitions occur on similar timescales. Using the Marcus-Levich-Jortner
type of electron transfer as a model reaction, we show that such dissipation
can change reactivity by accelerating internal thermalization, thereby
suppressing nonequilibrium effects that occur in the reaction outside
the cavity. This phenomenon is attributed mainly to cavity decay (i.e.,
photon leakage), but a supporting role is played by the relaxation
between polaritons and dark states. When nonequilibrium effects are
already suppressed in the bare reaction (the reactive species are
essentially at internal thermal equilibrium throughout the reaction),
we find that reactivity does not change significantly under VSC. Connections
are made between our results and experimental observations. 
\end{abstract}
\maketitle

\section{\label{sec:Introduction}Introduction}

Strong light-matter interaction, or simply strong coupling (SC), occurs
when an optical cavity mode and a material excitation coherently exchange
energy faster than either species decays.\citep{KavokinBook,Torma2015,Bitton2019,Mueller2020}
This interaction results in light-matter eigenstates called polaritons.
In addition to inheriting both photonic and material properties, the
hybrid states have different energies than their constituents. 

These features make polaritons a promising platform for modifying
the physicochemical properties of molecular systems.\citep{Ribeiro2018rev,Flick2018review}
Typically, SC in such systems is achieved by the interaction between
a cavity mode and $N\gg1$ quasidegenerate excitations in a molecular
ensemble. This collective SC produces two polariton states and $N-1$
dark states. At light-matter resonance, the so-called lower and upper
polaritons are respectively redshifted and blueshifted from the bare
molecular (photonic) excitations by the collective light-matter coupling
strength $g\sqrt{N}$, where $g$ is the single-molecule light-matter
coupling strength and $\Omega=2g\sqrt{N}$ is the so-called Rabi splitting.
In contrast, the dark states remain unchanged in energy. Nevertheless,
the polaritons can dissipate into the dark states,\citep{Agranovich2003,Virgili2011,delPino2015,Neuman2018optica,Xiang2018,Ribeiro2018vp,Scholes2020}
enriching the relaxation dynamics of the SC regime.

In light of the considerable impact that polariton formation has on
molecular excited states,\citep{Galego2015,Kowalewski2016,Szidarovszky2018,Haugland2020}
and of the photonic character of polaritons, it is not surprising
that SC is emerging as a versatile tool for manipulating photochemistry.
It has been shown that, in organic materials, polaritons can lower
the activation barrier of spin-conversion processes,\citep{Martinez-Martinez2018sf,Stranius2018,Martinez-Martinez2019,Yu2020chemRxiv,Ye2020ChemRxiv}
enhance conductivity,\citep{Orgiu2015,Schachenmayer2015,Feist2015,Liu2019,Krainova2020}
change the dynamics of photoinduced charge transfer,\citep{Herrera2016,Groenhof2018,Wang2020ChemRxiv,Mandal2020,Hoffmann2020,Mauro2020arXiv}
and alter photoisomerization yield.\citep{Hutchison2012,Galego2016,Fregoni2018,Mandal2019,Fregoni2020,Antoniou2020}
In some situations, dissipation associated with SC plays a key role
in the modification of photochemical processes. Notably, cavity decay
(photon leakage from the cavity) can enable the suppression of photodegradation
by diverting population from polaritons to the electronic ground state
instead of to the product.\citep{Munkhbat2018,Felicetti2020,Nefedkin2020}
The effect of cavity decay on photochemical reactivity has also been
demonstrated theoretically for photodissociation\citep{Ulusoy2020,Davidsson2020arxiv}
and photoisomerization\citep{Fregoni2020,Antoniou2020}. Another contributor
to the suppression of photodegradation is the relaxation between polaritons
and dark states.\citep{Munkhbat2018,Nefedkin2020} This dissipative
coupling can even mediate polariton-assisted remote energy transfer.\citep{Coles2014,Zhong2017,Du2018,Saez-Blazquez2018,Xiang2020}
Moreover, relaxation from polaritons to dark states can be harnessed to realize remote control of photoisomerization.\citep{Du2019}

Much less is understood about the ability to modify thermally activated
reactions using SC\citep{Shalabney2015,Casey2016} of molecular vibrations,
commonly known as vibrational SC (VSC). Recent experiments\citep{Hirai2020rev}
demonstrate that reactions can be enhanced or suppressed by VSC without
external pumping (e.g., laser excitation). Reactions affected by VSC
include organic substitution,\citep{Thomas2016,Thomas2019,Lather2019}
organic rearrangement,\citep{Hirai2020} and hydrolysis.\citep{Hiura2019ChemRxiv}
However, the intriguing VSC reaction kinetics still lacks a theoretical
underpinning. Transition-state theory, the most commonly used reaction-rate
theory, has been unsuccessful at explaining the VSC reactions.\citep{Galego2019,Campos-Gonzalez-Angulo2020,Zhdanov2020,Li2020}
While a model of nonadiabatic electron transfer under VSC captures
some of the observed trends,\citep{Campos-Gonzalez-Angulo2019} additional
connections to experiments have yet to be made. Common to the attempted
theoretical approaches is the following assumption: internal thermalization
(i.e., within the states of each chemical species) is much faster
than reactive transitions (i.e., between the states of different chemical
species). Said differently, states of each chemical species are assumed
to remain at internal thermal equilibrium throughout the reaction. 

Nonequilibrium effects may therefore be relevant in thermally activated
reactions modified by VSC. In the context of adiabatic reactions,
recent findings\citep{Li2020ChemRxiv} reveal that non-Markovian dynamics
along the cavity-mode coordinate can induce trapping of population
in a high-energy region of the potential energy surface, preventing
internal thermalization and promoting backward reactive transitions.
Alternatively, it would be interesting to explore how VSC affects
reactions with significant nonequilibrium effects outside the cavity.
For instance, some organic reactions involve the formation of vibrationally
hot intermediate species that react before they fully thermalize.\citep{Oyola2009,Glowacki2010,Carpenter2013rev}
This may happen in desilylation reactions, which feature reactive
intermediates\citep{Climent2020} and appear in studies reporting
suppression of product formation\citep{Thomas2016,Thomas2019,Thomas2020}
and rise in product selectivity\citep{Thomas2019} using VSC. If nonequilibrium
effects are indeed relevant to the VSC reactions, then another intriguing
topic is the role of dissipation, which is what enables thermal equilibrium
to be reached. 

Here, we carry out a kinetic study of how the dissipative processes
that VSC introduces to the chemical system---specifically, cavity
decay plus incoherent energy exchange among polaritons and dark states---affects
thermally activated reactions with significant nonequilibrium effects.
The class of reactions we consider is electron transfer, which is
modeled using Marcus-Levich-Jortner (MLJ) theory.\citep{Marcus1964,Levich1966,Jortner1976}
We examine reactions where vibrational relaxation and reactive transitions
occur on similar timescales. For the case of VSC, we use a generalization
of the MLJ model to calculate rates of reactive transitions. Rates
associated with internal thermalization are calculated using a standard
treatment\citep{delPino2015} of polariton relaxation. By comparing
reaction kinetics under VSC with those in the bare case and under
weak light-matter coupling, we show that dissipation associated with
VSC can alter reaction kinetics by suppressing nonequilibrium effects.
It follows that the presence of nonequilibrium effects in the bare
reaction can be important in determining the influence of VSC on reactivity.
Before concluding this paper, we discuss our results in the context
of the aforementioned experiments.

\section{\label{sec:Theory}Theory}

\subsection{\label{subsec:Hamiltonian}Hamiltonian}

Consider $N$ identical molecules inside an optical cavity. Each molecule
can undergo a series of nonadiabatic electron transfer reactions.
Such reactions occur via transitions between diabatic electronic states,
each representing a different reactive species (e.g., reactant, intermediate,
product). In the spirit of MLJ theory, the electronic states experience
vibronic coupling to high-frequency intramolecular vibrational modes
and low-frequency solvent vibrational modes,\citep{Jortner1976}
i.e., electron transfer is coupled to high- and low-frequency vibrations.
Both
types of modes are hence reactive degrees of freedom.
For simplicity,
suppose that each molecule has only one high-frequency reactive mode
(hereafter conveniently referred to as vibrational mode), and assume
that VSC takes place between this mode and a single cavity mode. Already,
we can see that the role of VSC is to modify the dynamics of a reactive
mode.

The Hamiltonian describing the electron-transfer molecules under VSC
is 
\begin{equation}
H=H_{\text{S}}+H_{\text{B}}+H_{\text{S}-\text{B}}.
\end{equation}
The first term, 
\begin{equation}
H_{\text{S}}=H_{\text{e}}+H_{\text{v}}+H_{\text{e}-\text{v}}+H_{\text{c}}+H_{\text{c}-\text{v}}+V_{\text{ET}},\label{eq:H_S}
\end{equation}
characterizes the ``system'', the subspace whose evolution we are
interested in. Comprising the system are the electronic, vibrational,
and cavity degrees of freedom. Electronic states are given by
\begin{equation}
H_{\text{e}}=\sum_{i=1}^{N}\sum_{\phi}E_{\phi}|\phi^{(i)}\rangle\langle\phi^{(i)}|,
\end{equation}
where $|\phi^{(i)}\rangle$ has energy $E_{\phi}$ and represents
the molecule $i$ belonging to reactive species $\phi$. Next, 
\begin{equation}
H_{\text{v}}=\hbar\omega_{\text{v}}\sum_{i=1}^{N}a_{i}^{\dagger}a_{i}\label{eq:H_v}
\end{equation}
characterizes the vibrational modes, where the mode belonging to molecule
$i$ is represented by creation (annihilation) operator $a_{i}^{\dagger}$
($a_{i}$). Vibronic coupling involving the high-frequency vibrational
modes is described by
\begin{equation}
H_{\text{e}-\text{v}}=\hbar\omega_{\text{v}}\sum_{i=1}^{N}\sum_{\phi}|\phi^{(i)}\rangle\langle\phi^{(i)}|\left[\lambda_{\phi}\left(a_{i}^{\dagger}+a_{i}\right)+\lambda_{\phi}^{2}\right].
\end{equation}
For any molecule of reactive species $\phi$, $\lambda_{\phi}$ is
the dimensionless displacement of the corresponding vibrational mode.
The following two terms of $H_{\text{S}}$ {[}Eq. (\ref{eq:H_S}){]} are
the cavity Hamiltonian
\begin{equation}
H_{\text{c}}=\hbar\omega_{\text{c}}a_{0}^{\dagger}a_{0},\label{eq:H_c}
\end{equation}
and the light-matter interaction
\begin{equation}
H_{\text{c}-\text{v}}=\hbar g\sum_{i=1}^{N}\left(a_{i}^{\dagger}a_{0}+
a_{0}^{\dagger}a_{i}
\right).\label{eq:H_c-v}
\end{equation}
The cavity mode has frequency $\omega_{\text{c}}$ and is represented
by creation (annihilation) operator $a_{0}^{\dagger}$ ($a_{0}$){\color{red}.}
In writing Eq. (\ref{eq:H_c-v}), 
we have assumed that molecules of all reactive species undergo VSC.
Presumably, this condition has been realized in studies showing that 
organic desilylation\citep{Thomas2019} and TPPK-acetone cyclization\citep{Hiura2019ChemRxiv} 
can be modified by VSC, 
namely that involving a vibrational mode which is optically bright and nearly identical in energy 
for both reactant and product species.
For cases where only one of reactant and product experiences VSC,
theoretical modeling\citep{Campos-Gonzalez-Angulo2019,Phuc2020}  
predicts interesting effects such as
autocatalysis;\citep{Campos-Gonzalez-Angulo2019}
we do not focus here on situations where different reactive species couple unequally to the cavity. 
Finally, the electron-transfer interaction reads
\begin{equation}
V_{\text{ET}}=\sum_{i=1}^{N}\sum_{\phi\neq\varphi}\left(J_{\phi\varphi}|\varphi^{(i)}\rangle\langle\phi^{(i)}|+J_{\varphi\phi}|\phi^{(i)}\rangle\langle\varphi^{(i)}|\right),
\end{equation}
where $J_{\phi\varphi}=J_{\varphi\phi}$ is the diabatic coupling
between reactive species $\phi$ and $\varphi$. The remaining degrees
of freedom, which constitute the ``bath'', are captured in $H_{\text{B}}$.
The bath includes the low-frequency solvent modes that participate
in electron transfer, as well as modes that induce decay and incoherent
energy exchange among the polaritons and dark states. The system-bath
coupling responsible for these relaxation processes is characterized
by $H_{\text{S}-\text{B}}$. For the sake of brevity, we will not
explicitly define $H_{\text{B}}$ and $H_{\text{S}-\text{B}}$ here.

The eigenstates of the system, treating the electron-transfer interaction
($V_{\text{ET}}$) as perturbative, can be found by diagonalizing
the zeroth-order Hamiltonian
\begin{equation}
H_{0}=H_{\text{e}}+H_{\text{e}-\text{v}}+H_{\text{CV}},\label{eq:H_0}
\end{equation}
where $H_{\text{CV}}=H_{\text{v}}+H_{\text{c}}+H_{\text{c}-\text{v}}$
governs the subsystem that includes cavity and vibrational modes.
Two polariton and $N-1$ dark modes make up the eigenmodes of $H_{\text{CV}}$.
Let operators $\alpha_{q}=\sum_{i=0}^{N}c_{qi}a_{i}$ for $q=\pm,2,\dots,N$
represent the eigenmodes. The polariton modes respectively have frequencies
$\omega_{\pm}=\frac{1}{2}\left(\omega_{\text{c}}+\omega_{\text{v}}\pm\sqrt{(\omega_{\text{c}}-\omega_{\text{v}})^{2}+4g^{2}N}\right)$
and are given by\begin{subequations}
\begin{align}
\alpha_{+} & =(\cos\theta)a_{0}+(\sin\theta)\frac{1}{\sqrt{N}}\sum_{i=1}^{N}a_{i},\\
\alpha_{-} & =(\sin\theta)a_{0}-(\cos\theta)\frac{1}{\sqrt{N}}\sum_{i=1}^{N}a_{i},
\end{align}
\label{eq:pol}\end{subequations}where $\theta=\frac{1}{2}\tan^{-1}[2g\sqrt{N}/(\omega_{\text{c}}-\omega_{\text{v}})]$.
The dark modes all have frequency $\omega_{\text{v}}$, and each one
is represented by $\alpha_{q}$ for $q=2,\dots,N$. Due to the degeneracy
of the dark states, there exist multiple ways to express the dark
modes in terms of the bare modes (e.g., see Refs. \citep{Campos-Gonzalez-Angulo2019}
and \citep{Strashko2016}). We now define multiparticle states for
the various system degrees of freedom. Define $|\bm{\phi}\rangle\equiv\left|\phi_{1},\phi_{2},\dots,\phi_{N}\right\rangle $
as the multiparticle electronic state where molecule $i$ belongs
to reactive species $\phi_{i}$. For the cavity-vibrational modes,
we define $|\mathbf{m}\rangle\equiv|m_{+},m_{-},m_{2},\dots,m_{N}\rangle$,
which is an eigenstate of $H_{\text{CV}}$ with $m_{q}$ excitations
in mode $q$ and with energy $\sum_{q=\pm,2,\dots,N}m_{q}\hbar\omega_{q}$.
Moving to this multiparticle representation and carrying out some
additional rearrangements, we can write the zeroth-order system Hamiltonian
{[}Eq. (\ref{eq:H_0}){]} in the diagonal form
\begin{align}
H_{0} & =\sum_{\bm{\phi}}\sum_{\mathbf{m}}E_{\bm{\phi};\mathbf{m}}|\bm{\phi};\mathbf{m}\rangle\langle\bm{\phi};\mathbf{m}|.
\end{align}
The eigenstates
\begin{equation}
|\bm{\phi};\mathbf{m}\rangle\equiv|\bm{\phi}\rangle\otimes|\tilde{\mathbf{m}}_{(\bm{\phi})}\rangle\label{eq:phi-m}
\end{equation}
are expressed as products of an electronic state and a displaced cavity-vibrational
state. The latter is given by
\begin{align}
|\tilde{\mathbf{m}}_{(\bm{\phi})}\rangle & \equiv\left[\prod_{q=\pm,2,\dots,N}\mathcal{D}_{q}^{\dagger}(\lambda_{\bm{\phi}q})\right]|\mathbf{m}\rangle,\label{eq:m-tilde}
\end{align}
where $\mathcal{D}_{q}(\lambda)=\exp\left(\lambda\alpha_{q}^{\dagger}-\lambda^{*}\alpha_{q}\right)$
is the displacement operator for eigenmode $q$. The displacement
of mode $q$, when the system is in electronic state $|\bm{\phi}\rangle$,
is
\begin{equation}
\lambda_{\bm{\phi}q}\equiv\sum_{i=1}^{N}\lambda_{\phi_{i}q}^{(i)},\label{eq:lambda_phi-q}
\end{equation}
where
\begin{equation}
\lambda_{\phi_{i}q}^{(i)}\equiv c_{qi}\frac{\omega_{\text{v}}}{\omega_{q}}\lambda_{\phi_{i}}\label{eq:lambda_phi_i-q}
\end{equation}
is the contribution due to the vibronic coupling in molecule $i$
(when this molecule belongs to reactive species $\phi_{i}$). From
Eqs. (\ref{eq:lambda_phi-q})-(\ref{eq:lambda_phi_i-q}), it is evident
that VSC redistributes the displacements of the bare modes among the
polariton and dark modes. How displaced each eigenmode is depends
on its frequency and its overlap with the bare modes {[}Eq. (\ref{eq:lambda_phi_i-q}){]}.
Returning to the electronic-cavity-vibrational eigenstates $|\bm{\phi};\mathbf{m}\rangle$,
the corresponding energies are 
\begin{align}
E_{\bm{\phi};\mathbf{m}} & =\sum_{i=1}^{N}E_{\phi_{i}}+\sum_{q=\pm,2,\dots,N}m_{q}\hbar\omega_{q}\nonumber \\
 & \quad+\left(\hbar\omega_{\text{v}}\sum_{i=1}^{N}\lambda_{\phi_{i}}^{2}-\sum_{q=\pm,2,\dots,N}\hbar\omega_{q}\left|\lambda_{\bm{\phi}q}\right|^{2}\right).\label{eq:E_phi-m}
\end{align}
We see that VSC does not only change the energies of states containing
vibrational character {[}second summation in the first line of Eq.
(\ref{eq:E_phi-m}){]}, but it also induces an energy shift equal
to the difference in high-frequency reorganization energy before and
after VSC {[}second line of Eq. (\ref{eq:E_phi-m}){]}. 

\subsection{\label{subsec:Kinetic-model}Kinetic model}

Having obtained the system eigenstates and energies, we proceed to
discuss the kinetic model for simulating thermally activated nonadiabatic
electron transfer under VSC. We are interested in reactions where
nonequilibrium effects are significant outside the cavity. To focus
on how the reactions are impacted by VSC, including dissipation brought
about by VSC, we limit ourselves to the case of $N=2$ and 
neglect
multiply excited cavity-vibrational states;
the state-space truncation is a good approximation in
our
simulations (Sec. \ref{sec:Simulations}), where we choose parameters 
such that (1)
virtually 
all initial population resides in the zero- and first-excitation manifolds of the cavity-vibrational subspace
and (2) such that transitions within these lower manifolds are much faster than transitions into higher manifolds.
With these simplifications, we keep the dynamics
tractable enough for physical interpretation while still working in the
regime of collective VSC (i.e., $N>1$). In Sec. \ref{sec:Connection-to-experiments},
we discuss the case of many-molecule VSC. 

We now present the kinetic model. The evolution of the system is governed
by the master equation
\begin{align}
\frac{dp_{(\bm{\phi};\mathbf{m})}(t)}{dt} & =-\left[\sum_{(\bm{\phi}';\mathbf{m}')\neq(\bm{\phi};\mathbf{m})}k(\bm{\phi}';\mathbf{m}'|\bm{\phi};\mathbf{m})\right]p_{(\bm{\phi};\mathbf{m})}(t)\nonumber \\
 & \quad+\sum_{(\bm{\phi}';\mathbf{m}')\neq(\bm{\phi};\mathbf{m})}k(\bm{\phi};\mathbf{m}|\bm{\phi}';\mathbf{m}')p_{(\bm{\phi}';\mathbf{m}')}(t).\label{eq:master-eqn}
\end{align}
The quantity $p_{(\bm{\phi};\mathbf{m})}$ represents the population
of $|\bm{\phi};\mathbf{m}\rangle$, and $k(\bm{\phi}';\mathbf{m}'|\bm{\phi};\mathbf{m})$
denotes the rate of transition $|\bm{\phi};\mathbf{m}\rangle\rightarrow|\bm{\phi}';\mathbf{m}'\rangle$.
Processes of three types are included in the kinetic model: reactive
transitions; cavity-vibrational loss and gain; and relaxation among
polaritons and the dark state (there is only one dark state for $N=2$).
The latter two sets of processes occur within the same electronic
state and contribute to internal thermalization.

A reactive transition occurs between states $|\bm{\phi};\mathbf{m}\rangle$
and $|\bm{\phi}';\mathbf{m}'\rangle$ if the initial state is converted
to the final state by the electron-transfer reaction of molecule $i$
from reactive species $\varphi$ to reactive species $\varphi'\neq\varphi$
(i.e., $|\bm{\phi}\rangle\neq|\bm{\phi}'\rangle$, $\phi_{i}=\varphi$,
$\phi_{i}'=\varphi'$, $\phi_{j}=\phi_{j}'$ for $j\neq i$). The
corresponding transition rate, derived in analogy to the MLJ electron-transfer
rate,\citep{MayBook} is
\begin{align}
k(\bm{\phi}';\mathbf{m}'|\bm{\phi};\mathbf{m}) & =\sqrt{\frac{\pi}{\lambda_{\text{s}}^{(\varphi\varphi')}k_{B}T}}\frac{|J_{\varphi\varphi'}|^{2}}{\hbar}\left|\langle\tilde{\mathbf{m}}_{(\bm{\phi}')}'|\tilde{\mathbf{m}}_{(\bm{\phi})}\rangle\right|^{2}\nonumber \\
 & \quad\times\exp\left[-\frac{\left(E_{\bm{\phi}';\mathbf{m}'}-E_{\bm{\phi};\mathbf{m}}+\lambda_{\text{s}}^{(\varphi\varphi')}\right)^{2}}{4\lambda_{\text{s}}^{(\varphi\varphi')}k_{B}T}\right],\label{eq:k_reactive}
\end{align}
where $\lambda_{\text{s}}^{(\varphi\varphi')}=\lambda_{\text{s}}^{(\varphi'\varphi)}$
is the low-frequency reorganization energy for the reaction between
species $\varphi$ and $\varphi'$, 
$k_{B}$ is the Boltzmann constant, 
and $T$ is the temperature. 
The rate depends on a generalized
Franck-Condon factor, $\langle\tilde{\mathbf{m}}_{(\bm{\phi}')}'|\tilde{\mathbf{m}}_{(\bm{\phi})}\rangle$, 
for the cavity-vibrational states, where
\begin{equation}
\left|\langle\tilde{\mathbf{m}}_{(\bm{\phi}')}'|\tilde{\mathbf{m}}_{(\bm{\phi})}\rangle\right|^{2}=\left|\prod_{q=\pm,\text{d}}\langle m_{q}'|\mathcal{D}_{q}\left(\lambda_{\varphi'q}^{(i)}-\lambda_{\varphi q}^{(i)}\right)|m_{q}\rangle\right|^{2}.\label{eq:franck-condon}
\end{equation}
We have relabeled the single dark mode of the two-molecule system
as $q=\text{d}$. The undisplaced cavity-vibrational state $|m_{q}\rangle$
is the single-particle eigenstate of $H_{\text{CV}}$ with $m_{q}$
excitations in mode $q$. Matrix elements of the displacement operator
$\mathcal{D}_{q}(\lambda)$ with respect to the undisplaced cavity-vibrational
states are evaluated according to the property\citep{AgarwalBook}
\begin{align}
\langle m_{q}'|\mathcal{D}_{q}(\lambda)|m_{q}\rangle & =\sqrt{\frac{m_{q}!}{(m'_{q})!}}e^{-|\lambda|^{2}/2}\lambda^{m'_{q}-m_{q}}L_{m_{q}}^{m_{q}'-m_{q}}(|\lambda|^{2})\nonumber \\
 & \quad\text{for }m_{q}'\geq m_{q},\label{eq:D-Laguerre}
\end{align}
where $L_{n}^{k}(x)$ is an associated Laguerre polynomial; for $m_{q}'<m_{q}$,
the matrix element is evaluated using the same expression except with
$m_{q}'\leftrightarrow m_{q}$ and $\lambda\rightarrow-\lambda^{*}$. 

We next discuss cavity-vibrational loss and gain. First consider loss.
For the optical cavities used in previous experiments of VSC reactions,
bare cavity states decay via the leakage of photons out into free
space. In contrast, bare vibrational states decay through dissipation
into solvent and intermolecular modes. In VSC, the cavity-vibrational
states can decay through a combination of cavity and vibrational channels.
The decay of $|\bm{\phi},\hat{\mathbf{e}}_{q}\rangle$, where $\hat{\mathbf{e}}_{q}$
is the unit vector representing a single excitation of mode $q=\pm,\text{d}$,
has rate\citep{delPino2015,Martinez-Martinez2018sf} 
\begin{equation}
k(\bm{\phi};\mathbf{0}|\bm{\phi};\hat{\mathbf{e}}_{q})=|c_{q0}|^{2}\kappa+\left(\sum_{i=1}^{2}|c_{qi}|^{2}\right)\gamma.\label{eq:k_loss}
\end{equation}
The bare cavity decay rate is $\kappa$, and the bare vibrational
decay rate is $\gamma$. For nonzero temperatures, the reverse process,
gain, can occur. The corresponding rate is determined by detailed
balance: $k(\bm{\phi};\hat{\mathbf{e}}_{q}|\bm{\phi};\mathbf{0})=k(\bm{\phi};\mathbf{0}|\bm{\phi};\hat{\mathbf{e}}_{q})\exp(-\hbar\omega_{q}/k_{B}T)$. 

Lastly, we have relaxation among polaritons and the dark state. Processes
of this type originate from anharmonic coupling between vibrational
states and low-frequency molecular modes of the local environment.\citep{delPino2015,Dunkelberger2016,Ribeiro2018vp,Xiang2018}
The relaxation from $|\bm{\phi},\hat{\mathbf{e}}_{q}\rangle$ to $|\bm{\phi},\hat{\mathbf{e}}_{q'}\rangle$
(where $q,q'=\pm,\text{d}$ and $q\neq q'$) has rate\citep{delPino2015,Martinez-Martinez2018sf}
\begin{align}
k(\bm{\phi};\hat{\mathbf{e}}_{q'}|\bm{\phi};\hat{\mathbf{e}}_{q}) & =2\pi\left(\sum_{i=1}^{2}|c_{q'i}|^{2}|c_{qi}|^{2}\right)\nonumber \\
 & \quad\times\left\{ \Theta(-\omega)\left[\bar{n}(-\omega)+1\right]\mathcal{J}(-\omega)\right.\nonumber \\
 & \quad+\left.\Theta(\omega)\bar{n}(\omega)\mathcal{J}(\omega)\right\} ,\label{eq:k_exchange}
\end{align}
where $\Theta(\omega)$ is the Heaviside step function, $\bar{n}(\omega)=\left[\exp(\hbar\omega/k_{B}T)-1\right]^{-1}$
is the Bose-Einstein distribution function, and $\mathcal{J}(\omega)$
is the spectral density of the anharmonically coupled low-frequency
bath modes. For the simulations in Sec. \ref{sec:Simulations}, we
assume an Ohmic spectral density:\citep{delPino2015} $\mathcal{J}(\omega)=\eta\omega\exp\left[-(\omega/\omega_{\text{cut}})^{2}\right]$,
where $\eta$ is a dimensionless parameter representing the anharmonic
system-bath interaction, and $\omega_{\text{cut}}$ is the cutoff
frequency of the low-frequency bath modes. 

Now that we have presented the various processes captured by our kinetic
model, we note, for completeness, that $k(\bm{\phi}';\mathbf{m}'|\bm{\phi};\mathbf{m})=0$
for any transition $|\bm{\phi};\mathbf{m}\rangle\rightarrow|\bm{\phi}';\mathbf{m}'\rangle$
not induced by these processes, i.e., not described by Eqs. (\ref{eq:k_reactive}),
(\ref{eq:k_loss}), and (\ref{eq:k_exchange}). 

To conclude this section, we summarize how VSC influences reactive
transitions and internal thermalization within our kinetic model.
We do so in the context of reactions with nonequilibrium effects,
which arise when internal thermalization is not fast compared to reactive
transitions. First, recall that VSC alters energies and redistributes
the vibronic coupling of localized bare modes among the delocalized
eigenmodes (Sec. \ref{subsec:Hamiltonian}). In view of this, Eq.
(\ref{eq:k_reactive}) says that VSC affects the rate of reactive
transitions by changing activation energy---which is related to the
energies of the initial and final states---and the Franck-Condon
factor between initial and final states. These rate changes can lead
to modified reactivity when reactive species are at internal thermal
equilibrium and when they are not. On the other hand, the additional
relaxation channels created by VSC, i.e., cavity loss and gain for
polaritons {[}Eq. (\ref{eq:k_loss}){]} and relaxation among polaritons
and the dark state {[}Eq. (\ref{eq:k_exchange}){]}, help thermalize
the vibrational states and only change populations when reactive species
are not at internal thermal equilibrium. Thus, if creating these additional relaxation channels is
the dominant effect of VSC, then nonequilibrium effects will be suppressed, 
as we will see in Sec. \ref{sec:Simulations}. 

\section{\label{sec:Simulations}Simulations}

In this section, we perform kinetic simulations of thermally activated
nonadiabatic electron transfer under VSC. We simulate a set of representative
reactions whose reactive transitions are comparable in timescale to
internal thermalization. As discussed in Sec. \ref{sec:Introduction},
we are interested in understanding how such reactions are affected
by VSC-induced dissipative processes. To this end, we choose reaction
parameters that highlight the impact of cavity decay, as well as relaxation
among polaritons and the dark state. In addition, parameters are chosen
such that the dynamics of states with more than one cavity-vibrational
excitation are insignificant compared to the dynamics of states with
one or zero of such excitation (Sec. \ref{subsec:Kinetic-model}).
Parameters specific to each reaction are listed in Table \ref{tab:rxns}.
Throughout the simulations, we use $\hbar\omega_{\text{v}}=2000$
cm$^{-1}$ and assume a resonant cavity mode (i.e., $\omega_{\text{c}}=\omega_{\text{v}}$).
We also fix the temperature at $T=298$ K. Unless otherwise stated,
we use the following parameters to characterize internal thermalization:
$\gamma=0.01$ ps$^{-1}$ , $\kappa=1$ ps$^{-1}$, $\eta=0.001$,
and $\omega_{\text{cut}}=0.1\omega_{\text{v}}$. We note that the
chosen cavity decay rate ($\kappa$) is much faster than the chosen
vibrational decay rate ($\gamma$) and that both rates are similar
to those found in VSC experiments.\citep{Xiang2018,Ribeiro2018vp,Xiang2020}
In calculations of reactions under VSC, we choose $g=(0.03\omega_{\text{v}})/\sqrt{2}$,
which yields a Rabi splitting of $\Omega=0.06\omega_{\text{v}}$.
\begin{table*}
\caption{\label{tab:rxns}Parameters for the reactions simulated in this work.}

\begin{ruledtabular}
\begin{tabular}{cccc}
Reaction & Figure & Reaction type & Parameters\footnote{For all reactions, $E_{A}=0$ and $\lambda_{A}=0$. All $J_{\phi\varphi}$
not listed here are equal to 0. All parameters, except $\lambda_\phi$ (dimensionless), have units $\hbar \omega_{\text{v}}$.} \tabularnewline
\hline 
1 & \ref{fig:fast-bwd} & $A\rightarrow B$ & $E_{B}=-0.6$, $\lambda_{B}=1.5$, $J_{AB}=0.01$, $\lambda_{\text{s}}^{(AB)}=0.08$\tabularnewline
2 & \ref{fig:thermo-unfav} & $A\rightarrow B$ & $E_{B}=0.95$, $\lambda_{B}=1$, $J_{AB}=0.002$, $\lambda_{\text{s}}^{(AB)}=0.05$\tabularnewline
3 & \ref{fig:int} & $A\rightarrow B\rightarrow C$ & $E_{B}=-1.05$, $E_{C}=-1.35$, $\lambda_{B}=1.5$, $\lambda_{C}=4.5$,
$J_{AB}=0.0003$, $J_{BC}=0.02$, $\lambda_{\text{s}}^{(AB)}=0.05$,
$\lambda_{\text{s}}^{(BC)}=0.3$\tabularnewline
\end{tabular}
\end{ruledtabular}

\end{table*}

For comparison, we also simulate the reactions in the bare case (see
Appendix \ref{appendix:Bare-case} for kinetic model) and in the case
of weak light-matter coupling (see Appendix \ref{appendix:Weak-light-matter-coupling}
for kinetic model and its derivation). In the latter, we set the light-matter
coupling to 1\% of the value used for VSC. For all simulations in
the weak-coupling regime, cavity decay is fast compared to the overall
reaction dynamics (which can have a different timescale than the reactive
transitions; see Sec. \ref{sec:Connection-to-experiments} for further
discussion). Therefore, the weak light-matter coupling effectively
induces relaxation between vibrational and cavity states (Appendix
\ref{appendix:Weak-light-matter-coupling}).\citep{Metzger2019,Pelton2015} 

In all simulations, the initial population is a thermal distribution
of reactant eigenstates. The distribution is determined by the Boltzmann
probability function {[}i.e., $p_{(\bm{\phi};\mathbf{m})}\propto\exp(-E_{\bm{\phi};\mathbf{m}}/k_{B}T)${]}
and normalized to 1. In accordance with our truncation of the cavity-vibrational
subspace (Sec. \ref{subsec:Kinetic-model}), the initial distribution
consists only of states with $\leq1$ excitation in this subspace.

After calculating the population evolution for each state in the system,
the total population $N_{\varphi}(t)$ of each reactive species $\varphi$
at time $t$ is computed according to 
\begin{equation}
N_{\varphi}(t)=\sum_{(\bm{\phi};\mathbf{m})}p_{(\bm{\phi};\mathbf{m})}(t)\langle\bm{\phi};\mathbf{m}|\mathcal{P}_{\varphi}|\bm{\phi};\mathbf{m}\rangle.
\end{equation}
The projection operator 
\begin{equation}
\mathcal{P}_{\varphi}=\sum_{i=1}^{2}|\varphi^{(i)}\rangle\langle\varphi^{(i)}|
\end{equation}
counts the number of molecules belonging to reactive species $\varphi$. 

\begin{figure}
\includegraphics{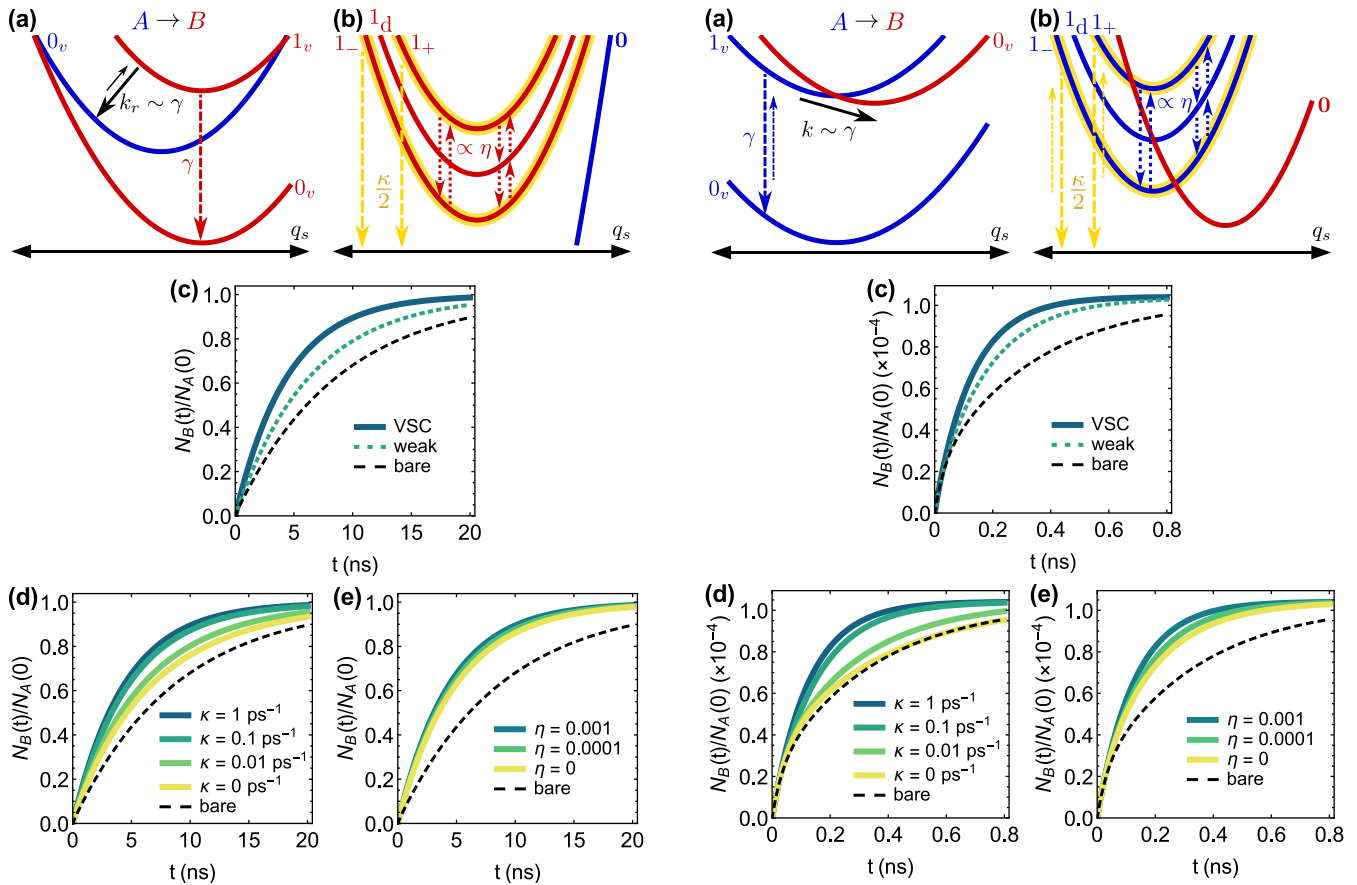}\caption{\label{fig:fast-bwd}(a) Scheme of reaction where a reverse reactive
transition (rate $k_r$) from the vibrationally hot product occurs on the same timescale
as vibrational decay (rate $\gamma$). 
Reactant (blue) and product (red) potential energy surfaces are plotted against effective solvent coordinate $q_s$.
Noteworthy transitions are shown. 
(Non)reactive transitions are represented with (dashed) solid arrows.
(b) Scheme depicting
VSC-induced internal thermalization of the product. 
Potential energy surfaces and nonreactive transitions,
shown here for the system under VSC, 
follow the formatting style used in (a),
with the additional attribute that polaritonic surfaces are drawn with yellow outline.
Polaritons can decay
via cavity leakage (rate $\kappa / 2$, 
where $\kappa$ is the decay rate of the bare cavity), 
and polaritons and dark state incoherently exchange
energy (rates $\propto \eta$, 
where $\eta$ represents the system-bath interaction
that gives rise to this relaxation). 
(c-e) Product population kinetics for various (c) regimes
of light-matter coupling, (d) $\kappa$, and (e)
$\eta$.}
\end{figure}

We first simulate a reaction where the vibrationally hot product undergoes
the reverse reaction as fast as it decays (Fig. \ref{fig:fast-bwd}a).
The main reactive transition involves a $0\rightarrow1$ vibrational
excitation upon going from reactant to product. The vibrationally
hot product either returns to the reactant via the reverse transition
or relaxes to the ground state, with both processes occurring at similar
rates. This scenario represents a nonequilibrium effect where the
product does not fully thermalize before it undergoes the reverse
reaction. Under VSC, the reaction is enhanced compared to the bare
case (Fig. \ref{fig:fast-bwd}c, blue solid line vs black dashed line).
The observed modification is consistent with the following mechanism:
dissipative processes associated with VSC speed up internal thermalization
and thereby force the product to decay instead of transforming back
into reactant. To test this mechanism, we carry out simulations for
various values of cavity decay rate $\kappa$ (Fig. \ref{fig:fast-bwd}d)
and anharmonic system-bath coupling parameter $\eta$ (Fig. \ref{fig:fast-bwd}e).
The latter determines the rates of relaxation among polaritons and
the dark state (Fig. \ref{fig:fast-bwd}b, red dotted arrows). As
cavity decay rate $\kappa$ is lowered to zero, most of the reaction
enhancement goes away (Figs. \ref{fig:fast-bwd}e). In contrast, the
modification largely survives as $\eta$ is decreased. These trends
reveal not only that the reaction is enhanced by accelerated internal
thermalization, but also that cavity decay plays the main role in
speeding up the thermalization. Specifically, cavity decay of the
polaritons (Fig. \ref{fig:fast-bwd}b, yellow dashed arrows) enables
the vibrationally hot product to cool much faster than it can undergo
a reverse reactive transition. Relaxation from the dark state to polaritons
(Fig. \ref{fig:fast-bwd}b, red dotted arrows) plays a supporting
role: it transfers population to the (polariton) states that decay
via cavity leakage. 

\begin{figure}
\includegraphics{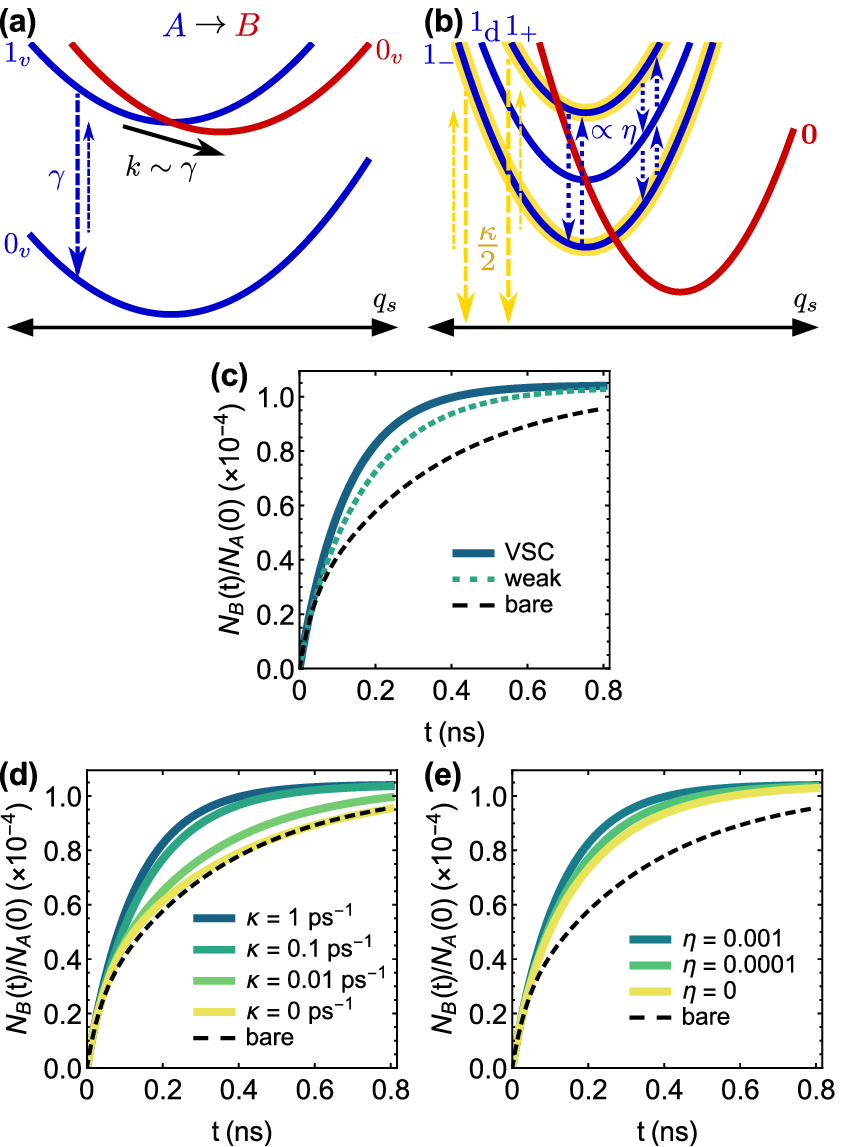}\caption{\label{fig:thermo-unfav} (a) Scheme of reaction where a reactive
transition (rate $k$) from a vibrational excited state occurs on the same timescale
as vibrational decay (rate $\gamma$). 
Reactant (blue) and product (red) potential energy surfaces are plotted against effective solvent coordinate $q_s$. 
Noteworthy transitions are shown. 
(Non)reactive transitions are represented with (dashed) solid arrows.
(b) Scheme depicting
VSC-induced internal thermalization of the reactant. 
Potential energy surfaces and nonreactive transitions,
shown here for the system under VSC, 
follow the formatting style used in (a),
with the additional attribute that polaritonic surfaces are drawn with yellow outline.
Polaritons 
can gain energy from the ground cavity-vibrational state via photon absorption
from the surroundings into the cavity, and polaritons and dark state
incoherently exchange energy. 
See caption of Fig. \ref{fig:fast-bwd} for explanation of $\kappa$, $\eta$, and labels containing these symbols.
(c-e) Product population kinetics for
various (c) regimes of light-matter coupling, 
(d) $\kappa$, 
and (e) $\eta$.}
\end{figure}

The second reaction we investigate has a reactive transition that
starts from a vibrational excited state and occurs on the same timescale
as vibrational decay (Fig. \ref{fig:thermo-unfav}a). Although thermodynamically
unfavorable, this reaction can serve as an instructional example for
thermodynamically favorable situations where the most reactive channel
involves a vibrational excited state in the reactant. Due to a large
energy difference between reactant and product electronic states,
there is only one reactive transition, which is accompanied by a $1\rightarrow0$
vibrational deexcitation. This transition happens at a rate comparable
to the decay of the initial vibrational state. Since vibrational decay
is extremely fast compared to its reverse process, the timescale of
reactant internal thermalization is that of vibrational decay, i.e.,
that of the reactive transition. Intuitively speaking, the population
of the reactive vibrational excited state does not reach its fully
replenished (thermalized) value prior to subsequent reactive transitions.
In the presence of VSC, the reaction experiences an enhancement (Fig.
\ref{fig:thermo-unfav}c, blue solid line vs black dashed line). By
varying $\kappa$ (Fig. \ref{fig:thermo-unfav}d) and $\eta$ (Fig.
\ref{fig:thermo-unfav}e), we again find that the reaction modification
is attributed mostly to cavity decay and less so to relaxation among
polaritons and the dark state. Cavity decay facilitates rapid thermalization
between the ground state and polaritons (Fig. \ref{fig:thermo-unfav}b,
yellow dashed arrows), while the dark state quickly reaches thermal
equilibrium through incoherent energy transfer with the polaritons
(Fig. \ref{fig:thermo-unfav}b, blue dotted arrows). As a result of
these relaxation dynamics, the excited cavity-vibrational states are
refilled with population before the next reactive transition takes
place. 

\begin{figure*}
\includegraphics{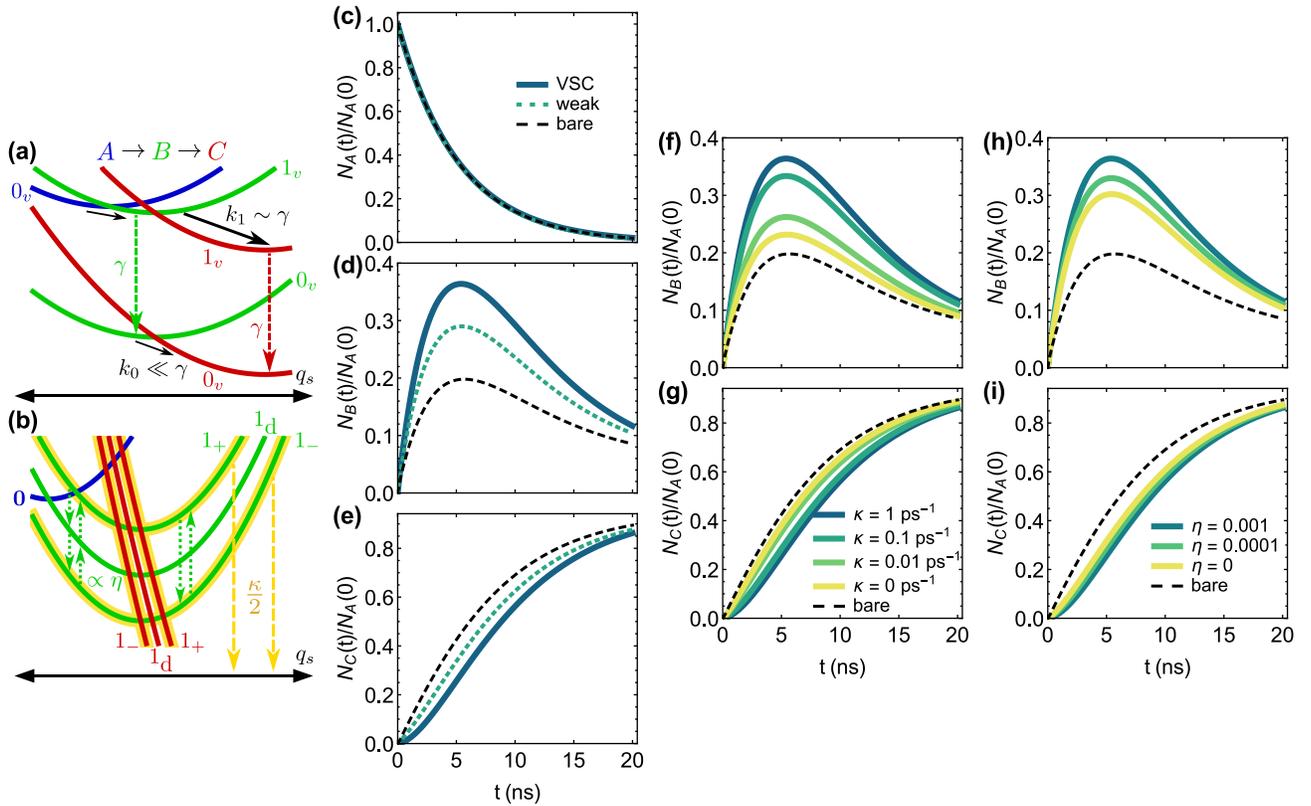}\caption{\label{fig:int} 
(a) Scheme of reaction where a reactive transition (rate $k_1$)
from a vibrationally hot intermediate occurs on the same timescale
as vibrational decay (rate $\gamma$). 
Reactant (blue), intermediate (green), and product (red) potential energy surfaces are plotted against effective solvent coordinate $q_s$.
Noteworthy transitions are shown. 
(Non)reactive transitions are represented with (dashed) solid arrows.
The rate $k_0$ describes a reactive transition from the intermediate in its vibrational ground state.
Note: $k_{0}\protect\neq k_{1}$
because the corresponding reactive transitions are associated with
different Franck-Condon factors {[}Eq. (\ref{eq:franck-condon_bare}){]}.
(b) Scheme depicting VSC-induced internal thermalization of the intermediate.
Potential energy surfaces and nonreactive transitions,
shown here for the system under VSC, 
follow the formatting style used in (a),
with the additional attribute that polaritonic surfaces are drawn with yellow outline.
Polaritons can decay via cavity leakage, and polaritons and dark
state incoherently exchange energy. 
See caption of Fig. \ref{fig:fast-bwd} for explanation of $\kappa$, $\eta$, and labels containing these symbols.
(c-e) Population kinetics of (c)
reactant, (d) intermediate, and (e) product for various regimes of
light-matter coupling. (f-i) Population kinetics of (f, h) intermediate
and (g, i) product for various (f, g) $\kappa$,
and (h, i) $\eta$.}
\end{figure*}

While the first two reactions are enhanced by VSC, suppression is
found for the third reaction, which involves a vibrationally hot intermediate
that reacts before it can fully thermalize (Fig. \ref{fig:int}a).\citep{Oyola2009,Glowacki2010,Carpenter2013rev}
The first reactive transition is from reactant to intermediate and
involves a $0\rightarrow1$ vibrational excitation. Next, the vibrationally
hot intermediate can either make a reactive transition to product
or relax to the ground state, with similar rates for both processes.
The former process is followed by vibrational decay, while the latter
is followed by an intermediate-to-product reactive transition that
is much slower than vibrational decay. Overall, a significant amount
of intermediate population first makes a fast reactive transition
and then cools, and the remaining intermediate population first cools
and then makes a slow reactive transition. With VSC, the population
kinetics of the reactant is unchanged (Fig. \ref{fig:int}c, blue
solid line vs black dashed line) because the vibrationally hot intermediate,
with or without VSC, is transformed into a different state much quicker
than it can transition back to reactant. However, there is greater
accumulation of the intermediate (Fig. \ref{fig:int}d, blue solid
line vs black dashed line) and suppressed formation of the product
under VSC (Fig. \ref{fig:int}e, blue solid line vs black dashed line).
Carrying out the same analysis as done with the previous two reactions
(see Figs. \ref{fig:int}f-\ref{fig:int}i), we conclude that the
rapid decay of polaritons via cavity leakage (Fig. \ref{fig:int}b,
yellow dashed arrows), assisted by the relaxation from the dark state
to polaritons (Fig. \ref{fig:int}b, green dotted arrows), causes
the intermediate to reach thermal equilibrium before it can go through
the faster reactive transition. 

To obtain additional insight regarding nonequilibrium effects and
how VSC alters reactivity, we study the reactions under weak light-matter
coupling. Figs. \ref{fig:fast-bwd}c, \ref{fig:thermo-unfav}c, \ref{fig:int}d-\ref{fig:int}e
show population kinetics for VSC (blue solid line), weak light-matter
coupling (aquamarine dotted line), and the bare case (black dashed
line). For all reactions, weak light-matter coupling leads to significantly
modified kinetics. The direction of change is the same as that of
VSC. This finding is consistent with the fact that, similar to VSC,
weak light-matter coupling opens up relaxation between vibrations
and the cavity, and enables internal thermalization of the reactive
species to be sped up by cavity decay. It also makes sense that the
magnitude of change is less than that of VSC, since the vibration-cavity
relaxation resulting from weak light-matter interaction is not as
fast as cavity decay.\citep{Metzger2019} Nevertheless, it is interesting
that one only needs to enter the weak-coupling regime to manipulate
reactivity. The same conclusion has been reached in a study of how
cavity decay allows light-matter coupling to protect molecules from
photodegradation.\citep{Felicetti2020} 

\begin{figure}
\includegraphics{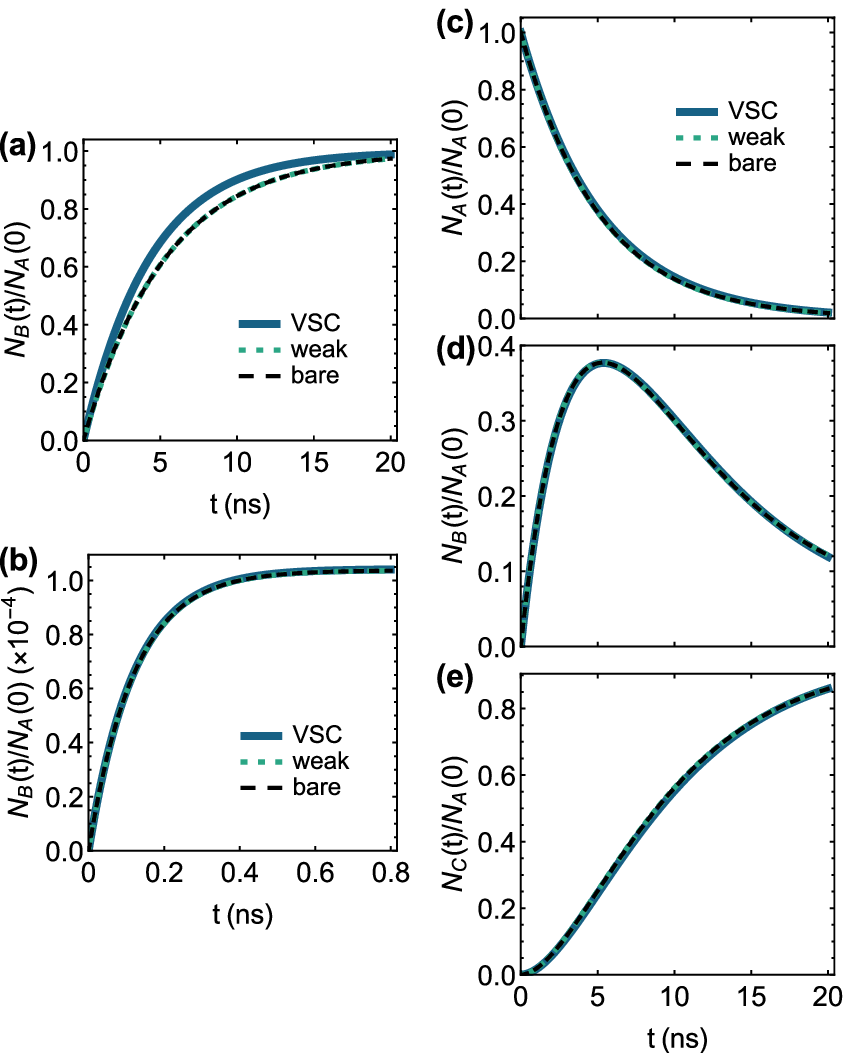}\caption{\label{fig:fast-vib-rel} Population kinetics with fast vibrational
decay rate $\gamma=1$ ps$^{-1}$. Plots correspond to the following
figures showing population kinetics with vibrational decay rate $\gamma=0.01$
ps$^{-1}$ but otherwise same conditions: (a) Fig. \ref{fig:fast-bwd}c,
(b) Fig. \ref{fig:thermo-unfav}c, and (c-e) Figs. \ref{fig:int}c--\ref{fig:int}e.}
\end{figure}

The simulations discussed so far suggest that the VSC-induced modifications
to the above reactions arise mainly from the suppression of nonequilibrium
effects. To strengthen this claim, we simulate the same reactions
except the vibrational decay rate is made 100 times larger (i.e.,
$\gamma=1\text{ ps}^{-1}=\kappa$). With this condition, we find that
the VSC and bare kinetics (Fig. \ref{fig:fast-vib-rel}, blue solid
line vs black dashed line) are either virtually identical or much
more similar than in the case of slow vibrational decay. Thus, the
changes in reactivity due to VSC are considerably reduced when, outside
the cavity, internal thermalization is already rapid compared to reactive
transitions. 

While our focus has been on nonequilibrium effects and their suppression
by VSC-related dissipation, we should note that the calculated reaction
kinetics do exhibit some changes as a consequence of VSC modifying
activation energies and redistributing vibronic coupling among the
cavity-vibrational eigenmodes (Sec. \ref{subsec:Kinetic-model}).
Because cavity decay has been identified as the major contributor
to the altered kinetics in the above reactions, the impact of the
minor contributors could be revealed by comparing simulations for
VSC and no cavity decay (i.e., $\kappa=0$ ps$^{-1}$) with those
for the bare case. For the reaction of Fig. \ref{fig:fast-bwd}, the
kinetics is noticeably different for the two conditions (Fig. \ref{fig:fast-bwd}d,
yellow solid line vs black dashed line). It seems that the difference
cannot be fully accounted for by the presence of relaxation among
polaritons and dark state (Fig. \ref{fig:fast-bwd}e). Hence, modified
energy differences and Franck-Condon factors likely impart a small
but appreciable influence on the reaction. This conclusion is corroborated
by the slight enhancement observed for the VSC reaction when vibrational
decay is made faster than all reactive transitions (Fig. \ref{fig:fast-vib-rel}a,
blue solid line vs black dashed line); in this regime, dissipative
processes introduced by VSC should have no effect. For further discussion
of how changes in energetics and Franck-Condon factors due to VSC
manifest as changes in reactivity, we refer the reader to Refs. \citep{Campos-Gonzalez-Angulo2019}
and \citep{Phuc2020} (see also Ref. \citep{Semenov2019}, which discusses
these topics but in the context of SC to electronic states).

\section{\label{sec:Connection-to-experiments}Connection to experiments}

A noteworthy difference between our simulations and the experiments
of VSC reactions is the number $N$ of molecules involved in VSC.
In our simulations $N=2$. In the experiments, $N$ is estimated to
be $10^{6}$ to $10^{12}$.\citep{Campos-Gonzalez-Angulo2019,delPino2015,Vurgaftman2020} 

For such large $N$, the particular reactions studied here are unlikely
to be modified by VSC. To understand this statement and to identify reactions
that would be significantly affected by many-molecule VSC, we present
the following argument. Without loss of generality, consider a reaction
featuring a single forward reactive transition, which involves a $0\rightarrow1$
vibrational excitation. This hypothetical reaction resembles that
depicted in Fig. \ref{fig:fast-bwd}a. Assuming the steady-state approximation
(SSA) for the vibrationally excited product and ignoring vibrational
gain (i.e., the reverse process of vibrational decay), the reaction
rate can be written as $k_{R\rightarrow P}=k_{f}[k_{d}/(k_{r}+k_{d})]$,
where $k_{f}$ is the rate of the forward reactive transition, $k_{r}$
is the rate of the reverse reactive transition, and $k_{d}$ is the
decay rate of the vibrationally excited product. The quantity in square
brackets, $k_{d}/(k_{r}+k_{d})$, is the efficiency with which the
vibrationally excited product decays rather than returns to reactant.
In the reaction with VSC, there are $N+1$ forward reactive transitions,
each involving the $0\rightarrow1$ excitation of a polariton or dark
state; there are two and $N-1$ such transitions, respectively. We
first focus on the two polaritonic forward reactive transitions. Recall
that the Franck-Condon factor for a 0-1 vibronic transition is proportional
to the displacement between initial and final vibrational states {[}Eq.
(\ref{eq:D-Laguerre}) with $m_{q}'=1$ and $m_{q}=0${]}. Since the
overlap between a polariton mode and each bare vibrational mode is
$O(N^{-1/2})$ {[}Eq. (\ref{eq:pol}){]}, the displacements---and
therefore the 0-1 Franck-Condon factors---associated with the polaritons
are $O(N^{-1/2})$ times the corresponding bare values {[}Eqs. (\ref{eq:lambda_phi-q})-(\ref{eq:lambda_phi_i-q}){]}.
Then the polaritonic reactive transitions have rates that scale as
$O(N^{-1})$ because the rate of a reactive transition depends on
the square of a Franck-Condon factor {[}Eq. (\ref{eq:k_reactive}){]}.
Suppose that cavity decay is much faster than all other transitions
from the polaritons, including reverse reactive transitions and relaxation
to dark states. This scenario, which has been observed for cavity-coupled
W(CO)$_{6}$ in nonpolar solvents,\citep{Xiang2019jpca} results in
a decay efficiency of 1 for product states with an excitation in a
polariton mode. By applying the SSA to such states and neglecting
cavity gain (i.e., the reverse process of cavity decay), we arrive
at the following effective rate for the polaritonic reactive transitions:
$k_{R\rightarrow P}^{(p)}=\epsilon k_{f}/N$, where $\epsilon$ is
a dimensionless constant that accounts for changes in the reaction
rate caused by the polaritons having modified energies. In contrast,
we assume that the reactive transitions involving the dark states
do not afford changes in the reaction rate, given that the dark states
have similar decay dynamics to the bare vibrations.\citep{Agranovich2003,delPino2015,Dunkelberger2016,Dunkelberger2018,Xiang2018,Ribeiro2018vp,Xiang2019jpca}
Furthermore, relaxation from dark states to polaritons should be negligible:
this dissipative process is mediated by local system-bath interactions
(see Sec. \ref{subsec:Kinetic-model}) and has rates scaling as $O(N^{-1})$,\citep{delPino2015,Du2018,Martinez-Martinez2018sf}
corresponding to the probability of each molecule in either polariton
{[}Eq. (\ref{eq:pol}){]}. So, in order for the reaction to be modified
by VSC, the reaction rate ($k_{R\rightarrow P}^{(p)}$) due to the
polaritonic reactive transitions should be, at least, comparable to
the bare reaction rate ($k_{R\rightarrow P}$), i.e., 
\begin{equation}
\frac{\epsilon}{N}\gtrsim\frac{k_{d}}{k_{r}+k_{d}}.\label{eq:criteria}
\end{equation}
With arbitrary reactions in mind, criterion (\ref{eq:criteria}) says
that for systems with experimentally relevant values of $N$, VSC
can change the kinetics of thermally activated reactions if
\begin{enumerate}
\item the polaritons give rise to $\epsilon\gg1$, which means that activation
energies are significantly reduced compared to the bare case (this
scenario is discussed in Ref. \citep{Campos-Gonzalez-Angulo2019}),
and/or 
\item in the bare case, reverse reactive transitions are extremely fast
compared to internal thermalization, i.e., populations are efficiently
trapped in the reactants.
\end{enumerate}
When at least one of these two conditions is satisfied, a large-$N$
generalization of the kinetic model presented in Sec. \ref{subsec:Kinetic-model}
should predict altered reactivity under VSC. Indeed, Ref. \citep{Campos-Gonzalez-Angulo2019}
theoretically demonstrates the ability for condition 1 to enable VSC
catalysis of ground-state reactions when $N=10^{7}$. While such a
demonstration has not been carried out for condition 2, Ref. \citep{Polak2020}
(see Fig. 6a of that work) has shown that condition 2 enables electronic
SC to boost the efficiency of harvesting triplet excitons into singlet
polaritons, ultimately leading to enhanced photoluminescence compared
to the bare organic material (in this case, the reverse reactive transition
is the very rapid fission of singlets into triplets). 

Despite the above and other differences between this theoretical work
and the experiments, the suppression of nonequilibrium effects is
a general phenomena and could be relevant to the mechanism by which
VSC modifies reaction kinetics in the latter. According to our calculations,
the suppression of nonequilibrium effects can lead to significant
increase or decrease in reactivity, depending on the specific properties
of the reaction. Both types of modifications have been found experimentally.\citep{Hirai2020rev}
The resemblance does not stop there. Recall the experimental studies
showing that the desilylation of PTA using TBAF is suppressed by VSC.\citep{Thomas2016,Thomas2020}
This reaction involves an intermediate species.\citep{Climent2020}
Based on kinetic measurements, the authors propose that VSC forces
the reaction to go through a new pathway, which involves a different
intermediate. Compared to the intermediate of the bare reaction, the
intermediate in the VSC reaction forms more easily but reacts less
readily. We would like to propose an alternative mechanism based on
our results. In the bare case, there could be nonequilibrium effects
that cause slow conversion of reactant to intermediate (similar to
the reaction of Fig. \ref{fig:thermo-unfav}) and fast conversion
of intermediate to product (like in the reaction of Fig. \ref{fig:int};
see Refs. \citep{Oyola2009}, \citep{Glowacki2010}, and \citep{Carpenter2013rev}).
Under VSC, the intermediate would be the same as in the bare case,
but nonequilibrium effects would be suppressed, thereby accelerating
the reactant-to-intermediate transformation and decelerating the intermediate-to-product
transformation. This mechanism is consistent with the experimentally
proposed one. Future works should test the possibility of VSC-induced
suppression of nonequilibrium effects in the desilylation and other
reactions. 

When evaluating this possibility, it is important to keep in mind
that nonequilibrium effects can affect reactivity even if internal
thermalization occurs on a much shorter timescale than the reaction.
Whether or not nonequilibrium effects are present is determined by
how fast internal thermalization is relative to reactive transitions,
which may be faster than the (overall) reaction. These guiding principles
are supported by our simulations. For the bare reactions shown in
Figs. \ref{fig:fast-bwd} and \ref{fig:int}, internal thermalization
occurs on a 100 ps timescale while the reaction occurs on a 10 ns
timescale (Figs. \ref{fig:fast-bwd}a, \ref{fig:fast-bwd}c, \ref{fig:int}a,
\ref{fig:int}c). Nevertheless, as already discussed in Sec. \ref{sec:Simulations},
the ability of VSC---in particular, the associated relaxation processes---to
change the reaction kinetics relies on having (in the bare case) reactive
transitions with similar rates as internal thermalization. The above
principles, when made specific to adiabatic reactions, can be stated
as follows:\citep{Leitner1997} regardless of the actual reaction
rate or that predicted by transition-state theory, the influence that
nonequilibrium effects have on reactivity depends on how the rate
of internal thermalization compares to the reactive flux 
(i.e., the rate of product formation per unit population) 
at the transition-state
barrier.

\section{\label{sec:Conclusions}Conclusions}

In this work, we study how reactions with significant nonequilibrium
effects are influenced by the dissipative channels that VSC introduces
to the chemical system. By using the MLJ formalism of electron transfer
as our reaction model, we present a kinetic framework that captures
reactive transitions, vibrational decay, cavity decay, and relaxation
among polaritons and dark states. We then simulate reactions where
internal thermalization and reactive transitions occur on the same
timescale. The considered reactions respectively exhibit three representative
nonequilibrium effects: vibrationally hot product transforming back
to reactant while thermalizing (Fig. \ref{fig:fast-bwd}), slow replenishing
of reactive vibrational excited state after a reactive transition
(Fig. \ref{fig:thermo-unfav}), and vibrationally hot intermediate
reacting before it can fully thermalize (Fig. \ref{fig:int}). Under
VSC, the first two reactions are enhanced whereas the last reaction
is suppressed. These modifications largely occur because nonequilibrium
effects are suppressed by VSC-induced dissipative processes. The suppression
of nonequilibrium effects is driven by cavity decay, which causes
polaritons to thermalize much faster than reactive transitions occur.
Relaxation between polaritons and dark states allows the latter to
indirectly experience accelerated thermalization due to cavity decay.
When we substantially increase the vibrational decay rate, i.e., make
internal thermalization much faster than reactive transitions (for the cavity-free case), 
the bare and VSC reactivities become much closer (Fig.
\ref{fig:fast-vib-rel}). Finally, we discuss our work in the context
of recent experiments,\citep{Hirai2020rev} which demonstrate that
thermally activated reactions can be enhanced or suppressed by the
collective VSC of a large number of molecules. We identify types of
reactions for which our theory would predict modified kinetics induced
by many-molecule VSC. We also highlight resemblances between our results
and experimental observations. These resemblances suggest that VSC-triggered
suppression of nonequilibrium effects could play a role in the experiments. 
\begin{acknowledgments}
\textcolor{black}{The development of the presented models by M.D.
and J.Y.-Z. was supported by NSF CAREER CHE 1654732. The comparison
of these models with previous rate theories for thermally activated
polariton chemistry by J.A.C.-G.-A. was supported by AFOSR award FA9550-18-1-0289.
}M.D. is also supported by a UCSD Roger Tsien Fellowship while J.A.C.-G.-A.
is also supported by a UC-MEXUS/CONACYT graduate fellowship (Reference
No. 235273/472318). M.D. thanks Luis Mart\'inez-Mart\'inez, Sindhana
Pannir-Sivajothi, and Kai Schwennicke for useful discussions. 
\end{acknowledgments}

\section*{Data Availability}

The data that support the findings of this study are available from
the corresponding author upon reasonable request.

\appendix

\section{\label{appendix:Bare-case}Bare case}

For the kinetic model of reactions in the bare case (i.e., light-matter
coupling strength $g=0$), we write the zeroth-order system eigenstates
{[}i.e., the eigenstates of $H_{0}$, Eq. (\ref{eq:H_0}){]} in the
basis of localized vibrational and cavity modes {[}see Eqs. (\ref{eq:H_v})
and (\ref{eq:H_c}), respectively{]} instead of the basis of the polariton
and dark modes. In the chosen basis, the eigenstates are still of
the form $|\bm{\phi};\mathbf{m}\rangle\equiv|\bm{\phi}\rangle\otimes|\tilde{\mathbf{m}}_{(\bm{\phi})}\rangle$
{[}Eq. (\ref{eq:phi-m}){]}, except that the displaced cavity-vibrational
states are given by
\begin{equation}
|\tilde{\mathbf{m}}_{(\bm{\phi})}\rangle=\left[\prod_{i=1}^{N}\mathcal{D}_{i}^{\dagger}(\lambda_{\phi_{i}})\right]|\mathbf{m}\rangle,
\end{equation}
where $|\mathbf{m}\rangle\equiv|m_{0},m_{1},,\dots,m_{N}\rangle$
represents the state where mode $i=0,1,,\dots,N$ has $m_{i}$ excitations,
and $\mathcal{D}_{i}(\lambda)=\exp\left(\lambda a_{i}^{\dagger}-\lambda^{*}a_{i}\right)$
is the displacement operator for mode $i$. The zeroth-order energy
of $|\bm{\phi};\mathbf{m}\rangle$ is 
\begin{equation}
E_{\bm{\phi};\mathbf{m}}=\sum_{i=1}^{N}E_{\phi_{i}}+m_{0}\hbar\omega_{\text{c}}+\hbar\omega_{\text{v}}\sum_{i=1}^{N}m_{i}.
\end{equation}

The bare system is evolved under the same assumptions (including $N=2$)
as in the case of VSC (Sec. \ref{subsec:Kinetic-model}). The master
equation of the bare dynamics takes the general form of Eq. (\ref{eq:master-eqn}).
For the reaction of molecule $i=1,2$ from reactive species $\varphi$
to reactive species $\varphi'\neq\varphi$, the reactive-transition
rate is calculated with Eq. (\ref{eq:k_reactive}), where the generalized
Franck-Condon factors are evaluated using 
\begin{equation}
\left|\langle\tilde{\mathbf{m}}_{(\bm{\phi}')}'|\tilde{\mathbf{m}}_{(\bm{\phi})}\rangle\right|^{2}=\left|\delta_{m_{0}'m_{0}}\langle m_{i}'|\mathcal{D}_{i}(\lambda_{\varphi'}-\lambda_{\varphi})|m_{i}\rangle\delta_{m_{j}'m_{j}}\right|^{2}\label{eq:franck-condon_bare}
\end{equation}
for $j=2\delta_{i1}+\delta_{i2}\neq i$ (here, $\delta_{ik}$ is the
Kronecker delta) and the analog of Eq. (\ref{eq:D-Laguerre}) with
$q\rightarrow i$. The undisplaced single-particle state $|m_{i}\rangle$
represents $m_{i}$ excitations in mode $i$. The decay transitions
have rates
\begin{align}
k(\bm{\phi};\mathbf{0}|\bm{\phi};\hat{\mathbf{e}}_{i}) & =\begin{cases}
\kappa, & i=0,\\
\gamma, & i\neq0.
\end{cases}
\end{align}
The reverse rates are governed by detailed balance: 
\begin{equation}
k(\bm{\phi};\hat{\mathbf{e}}_{i}|\bm{\phi};\mathbf{0})=\begin{cases}
k(\bm{\phi};\mathbf{0}|\bm{\phi};\hat{\mathbf{e}}_{i})\exp(-\hbar\omega_{\text{c}}/k_{B}T), & i=0,\\
k(\bm{\phi};\mathbf{0}|\bm{\phi};\hat{\mathbf{e}}_{i})\exp(-\hbar\omega_{\text{v}}/k_{B}T), & i\neq0.
\end{cases}
\end{equation}
Among the singly excited vibrational states, there are no transitions
of the same nature as the relaxation among polaritons and dark states,
i.e., $k(\bm{\phi};\hat{\mathbf{e}}_{i'}|\bm{\phi};\hat{\mathbf{e}}_{i})=0$.
As mentioned in Sec. \ref{subsec:Kinetic-model}, this incoherent
energy exchange is caused by local system-bath interactions, which
do not couple different local vibrational modes. All rates $k(\bm{\phi}';\mathbf{m}'|\bm{\phi};\mathbf{m})$
that have not been discussed in this section are taken to be 0. 

\section{\label{appendix:Weak-light-matter-coupling}Weak light-matter coupling}

In this section, we derive the kinetic model that we use to simulate
reactions where the light-matter interaction strength is $g=(3\times10^{-4})\omega_{\text{v}}/\sqrt{2}$
and the cavity decay rate is $\kappa=1\text{ps}^{-1}\gg g\sqrt{N}$ (where $N=2$). Together,
these parameters signify the regime of weak light-matter coupling.
In this regime, the light-matter coupling can be treated as a perturbation
to the system. Thus, the zeroth-order system eigenstates are those
of the bare case (Appendix \ref{appendix:Bare-case}). However, the
light-matter coupling does change the population dynamics of the system.
Below we follow a textbook approach (see Ref. \citep{MayBook}, pp.
103-106 and 413) to show that, when the cavity and/or vibrational
excitations decay on a timescale much shorter than that of the reaction
dynamics, the effect of weak light-matter coupling is to induce incoherent
energy exchange between the excitations. Essentially, the employed
approach is a perturbative correction to the non-Hermitian dynamics
of the bare case. 

Formally, the dynamics of the bare system can be given by the quantum
master equation\citep{MayBook}
\begin{equation}
\frac{d\rho(t)}{dt}=-i(\mathcal{L}_{0}-i\mathcal{R})\rho(t),\label{eq:QME}
\end{equation}
where $\rho(t)$ is the reduced density operator of the system, and
the superoperator $\mathcal{L}_{0}-i\mathcal{R}$ generates the bare
dynamics. The Liouvillian superoperator $\mathcal{L}_{0}(\cdot)=[H_{0},\cdot]/\hbar$
generates the coherent (Hermitian) dynamics of the bare system. Here,
$H_{0}$ is the zeroth-order system Hamiltonian without light-matter
interaction. The superoperator $\mathcal{R}$ represents system-bath
interaction and generates the incoherent (non-Hermitian) dynamics
of the system, i.e., all reaction and relaxation dynamics of the bare
case. If we act on Eq. (\ref{eq:QME}) with $\langle\bm{\phi};\mathbf{m}|$
from the left and $|\bm{\phi};\mathbf{m}\rangle$ from the right,
we recover the master equation governing the bare population dynamics
(see Appendix \ref{appendix:Bare-case}):
\begin{align}
\frac{dp_{(\bm{\phi};\mathbf{m})}(t)}{dt} & =-k_{(\bm{\phi};\mathbf{m})}^{(\text{out})}p_{(\bm{\phi};\mathbf{m})}(t)\nonumber \\
 & \quad+\sum_{(\bm{\phi}';\mathbf{m}')\neq(\bm{\phi};\mathbf{m})}k(\bm{\phi};\mathbf{m}|\bm{\phi}';\mathbf{m}')p_{(\bm{\phi}';\mathbf{m}')}(t),\label{eq:master-eqn_k^out}
\end{align}
where, for convenience, we have defined $k_{(\bm{\phi};\mathbf{m})}^{(\text{out})}\equiv\sum_{(\bm{\phi}';\mathbf{m}')\neq(\bm{\phi};\mathbf{m})}k(\bm{\phi}';\mathbf{m}'|\bm{\phi};\mathbf{m})$
as the outgoing rate of population transfer from state $|\bm{\phi};\mathbf{m}\rangle$.
If we instead act on Eq. (\ref{eq:QME}) with $\langle\bm{\phi};\mathbf{m}|$
from the left and $|\bm{\phi}';\mathbf{m}'\rangle\neq|\bm{\phi};\mathbf{m}\rangle$
from the right, we arrive at
\begin{align}
\frac{d\rho_{(\bm{\phi};\mathbf{m}),(\bm{\phi}';\mathbf{m}')}(t)}{dt} & =-i\Bigl[(E_{\bm{\phi};\mathbf{m}}-E_{\bm{\phi}';\mathbf{m}'})/\hbar\nonumber \\
 & \quad-ik_{(\bm{\phi};\mathbf{m})}^{(\text{out})}/2-ik_{(\bm{\phi}';\mathbf{m}')}^{(\text{out})}/2\Bigr]\rho_{(\bm{\phi};\mathbf{m}),(\bm{\phi}';\mathbf{m}')}(t),\label{eq:coh}
\end{align}
where $\rho_{(\bm{\phi};\mathbf{m}),(\bm{\phi}';\mathbf{m}')}(t)=\langle\bm{\phi};\mathbf{m}|\rho(t)|\bm{\phi}';\mathbf{m}'\rangle$
is a coherence. In writing Eq. (\ref{eq:coh}), we have assumed that
each coherence is decoupled from populations and other coherences.
We have also, for simplicity, neglected pure dephasing. 

For the system under weak light-matter coupling, the light-matter
interaction $H_{\text{c}-\text{v}}$ {[}Eq. (\ref{eq:H_c-v}){]} is
added to the Hamiltonian ($H_{0}$) of the bare system. To account
for this change in the quantum master equation, we simply add the
superoperator $\mathcal{L}_{\text{c}-\text{v}}(\cdot)=[H_{\text{c}-\text{v}},\cdot]/\hbar$
to the superoperator ($\mathcal{L}_{0}-i\mathcal{R}$) that generates
the bare dynamics. In doing so, we implicitly ignore the effect of
light-matter coupling on system-bath interaction; this approximation
should hold for sufficiently small values of light-matter interaction
strength. Quantum master equation (\ref{eq:QME}) now reads
\begin{equation}
\frac{d\rho(t)}{dt}=-i(\mathcal{L}_{0}-i\mathcal{R})\rho(t)-i\mathcal{L}_{\text{c}-\text{v}}\rho(t).
\end{equation}
With this transformation, states with a single vibrational excitation
(i.e., $|\bm{\phi},\hat{\mathbf{e}}_{i}\rangle$, $i=1,2$) evolve
as 
\begin{align}
\frac{dp_{(\bm{\phi};\hat{\mathbf{e}}_{i})}(t)}{dt} & =-k_{(\bm{\phi};\hat{\mathbf{e}}_{i})}^{(\text{out})}p_{(\bm{\phi};\hat{\mathbf{e}}_{i})}(t)\nonumber \\
 & \quad+\sum_{(\bm{\phi}';\mathbf{m}')\neq(\bm{\phi};\hat{\mathbf{e}}_{i})}k(\bm{\phi};\hat{\mathbf{e}}_{i}|\bm{\phi}';\mathbf{m}')p_{(\bm{\phi}';\mathbf{m}')}(t)\nonumber \\
 & \quad+2g\text{Im}\rho_{(\bm{\phi};\hat{\mathbf{e}}_{0}),(\bm{\phi};\hat{\mathbf{e}}_{i})}(t),\label{eq:new-vib}
\end{align}
and states with a single cavity excitation (i.e., $|\bm{\phi},\hat{\mathbf{e}}_{0}\rangle$)
evolve as 
\begin{align}
\frac{dp_{(\bm{\phi};\hat{\mathbf{e}}_{0})}(t)}{dt} & =-k_{(\bm{\phi};\hat{\mathbf{e}}_{0})}^{(\text{out})}p_{(\bm{\phi};\hat{\mathbf{e}}_{0})}(t)\nonumber \\
 & \quad+\sum_{(\bm{\phi}';\mathbf{m}')\neq(\bm{\phi};\hat{\mathbf{e}}_{0})}k(\bm{\phi};\hat{\mathbf{e}}_{0}|\bm{\phi}';\mathbf{m}')p_{(\bm{\phi}';\mathbf{m}')}(t)\nonumber \\
 & \quad-2g\text{Im}\sum_{j=1}^{2}\rho_{(\bm{\phi};\hat{\mathbf{e}}_{0}),(\bm{\phi};\hat{\mathbf{e}}_{j})}(t).\label{eq:new-cav}
\end{align}
To arrive at Eqs (\ref{eq:new-vib})-(\ref{eq:new-cav}), we have
used the general relation $\rho_{(\bm{\phi}';\mathbf{m}'),(\bm{\phi};\mathbf{m})}=\rho_{(\bm{\phi};\mathbf{m}),(\bm{\phi}';\mathbf{m}')}^{*}$.
For states with no vibrational or cavity excitations, the equations
of motion remain the same as in the bare case. From Eqs. (\ref{eq:new-vib})
and (\ref{eq:new-cav})---in particular, the last line of each equation---it
is clear that the evolution of vibrational and cavity populations
now depends on the light-matter coherence $\rho_{(\bm{\phi};\hat{\mathbf{e}}_{0}),(\bm{\phi};\hat{\mathbf{e}}_{i})}$.
In the presence of weak light-matter coupling, this light-matter coherence
evolves according to
\begin{align}
 & \frac{d\rho_{(\bm{\phi};\hat{\mathbf{e}}_{0}),(\bm{\phi};\hat{\mathbf{e}}_{i})}(t)}{dt}\nonumber \\
 & =-i\left(\Delta-ik_{(\bm{\phi};\hat{\mathbf{e}}_{0})}^{(\text{out})}/2-ik_{(\bm{\phi};\hat{\mathbf{e}}_{i})}^{(\text{out})}/2\right)\rho_{(\bm{\phi};\hat{\mathbf{e}}_{0}),(\bm{\phi};\hat{\mathbf{e}}_{i})}(t)\nonumber \\
 & \quad-igp_{(\bm{\phi};\hat{\mathbf{e}}_{i})}(t)+igp_{(\bm{\phi};\hat{\mathbf{e}}_{0})}(t)\nonumber \\
 & \quad-ig\rho_{(\bm{\phi};\hat{\mathbf{e}}_{j}),(\bm{\phi};\hat{\mathbf{e}}_{i})}(t),\label{eq:coh-LM-1}
\end{align}
where $j=2\delta_{i1}+\delta_{i2}\neq i$. We have defined $\Delta\equiv\omega_{\text{c}}-\omega_{\text{v}}$.
Notice the appearance of $\rho_{(\bm{\phi};\hat{\mathbf{e}}_{j}),(\bm{\phi};\hat{\mathbf{e}}_{i})}$,
a coherence involving different vibrational states, in the last line
of Eq. (\ref{eq:coh-LM-1}). In the weak-coupling regime, the evolution
of this purely vibrational coherence is given by
\begin{align}
\frac{d\rho_{(\bm{\phi};\hat{\mathbf{e}}_{j}),(\bm{\phi};\hat{\mathbf{e}}_{i})}(t)}{dt} & =-\left(k_{(\bm{\phi};\hat{\mathbf{e}}_{j})}^{(\text{out})}/2+k_{(\bm{\phi};\hat{\mathbf{e}}_{i})}^{(\text{out})}/2\right)\rho_{(\bm{\phi};\hat{\mathbf{e}}_{j}),(\bm{\phi};\hat{\mathbf{e}}_{i})}(t)\nonumber \\
 & \quad-ig\left[\rho_{(\bm{\phi};\hat{\mathbf{e}}_{0}),(\bm{\phi};\hat{\mathbf{e}}_{i})}(t)-\rho_{(\bm{\phi};\hat{\mathbf{e}}_{j}),(\bm{\phi};\hat{\mathbf{e}}_{0})}(t)\right].
\end{align}
It is apparent that the coherence $\rho_{(\bm{\phi};\hat{\mathbf{e}}_{j}),(\bm{\phi};\hat{\mathbf{e}}_{i})}$
does not directly couple to populations. In other words, the coupling
of populations to the purely vibrational coherence is higher-order
in $g$, the strength of the perturbative light-matter interaction.
We are interested in the lowest-order correction to the population
dynamics due to light-matter interaction. Hence, we neglect the contribution
of $\rho_{(\bm{\phi};\hat{\mathbf{e}}_{j}),(\bm{\phi};\hat{\mathbf{e}}_{i})}$
in Eq. (\ref{eq:coh-LM-1}). This truncation, followed by formal integration
of Eq. (\ref{eq:coh-LM-1}), leads to
\begin{align}
 & \rho_{(\bm{\phi};\hat{\mathbf{e}}_{0}),(\bm{\phi};\hat{\mathbf{e}}_{i})}(t)\nonumber \\
 & =-ig\int_{0}^{t}dt'\exp\left[-i\left(\Delta-ik_{(\bm{\phi};\hat{\mathbf{e}}_{0})}^{(\text{out})}/2-ik_{(\bm{\phi};\hat{\mathbf{e}}_{i})}^{(\text{out})}/2\right)t'\right]\nonumber \\
 & \quad\times\left[p_{(\bm{\phi};\hat{\mathbf{e}}_{i})}(t-t')-p_{(\bm{\phi};\hat{\mathbf{e}}_{0})}(t-t')\right],
\end{align}
where we have made the change of variable $(t-t')\rightarrow t'$.
For our simulations, we are interested in how populations change over
times much longer than cavity and vibrational decay. Since these two
processes are included in $k_{(\bm{\phi};\hat{\mathbf{e}}_{0})}^{(\text{out})}$
and $k_{(\bm{\phi};\hat{\mathbf{e}}_{i})}^{(\text{out})}$, respectively,
then the exponential term will decay on a timescale much shorter than
the timescales of interest. We therefore make a Markovian approximation---i.e.,
the substitutions $p_{(\bm{\phi};\hat{\mathbf{e}}_{i})}(t-t')\rightarrow p_{(\bm{\phi};\hat{\mathbf{e}}_{i})}(t)$
and $p_{(\bm{\phi};\hat{\mathbf{e}}_{0})}(t-t')\rightarrow p_{(\bm{\phi};\hat{\mathbf{e}}_{0})}(t)$,
as well as extending the integral to infinity---to obtain
\begin{align}
 & \rho_{(\bm{\phi};\hat{\mathbf{e}}_{0}),(\bm{\phi};\hat{\mathbf{e}}_{i})}(t)\nonumber \\
 & =-ig\left[p_{(\bm{\phi};\hat{\mathbf{e}}_{i})}(t)-p_{(\bm{\phi};\hat{\mathbf{e}}_{0})}(t)\right]\nonumber \\
 & \quad\times\int_{0}^{\infty}dt'\exp\left[-i\left(\Delta-ik_{(\bm{\phi};\hat{\mathbf{e}}_{0})}^{(\text{out})}/2-ik_{(\bm{\phi};\hat{\mathbf{e}}_{i})}^{(\text{out})}/2\right)t'\right].\label{eq:after-markov}
\end{align}
Evaluating the integral and plugging Eq. (\ref{eq:after-markov})
into Eqs. (\ref{eq:new-vib})-(\ref{eq:new-cav}), we arrive at
\begin{align}
\frac{dp_{(\bm{\phi};\hat{\mathbf{e}}_{i})}(t)}{dt} & =-k_{(\bm{\phi};\hat{\mathbf{e}}_{i})}^{(\text{out})}p_{(\bm{\phi};\hat{\mathbf{e}}_{i})}(t)\nonumber \\
 & \quad+\sum_{(\bm{\phi}';\mathbf{m}')\neq(\bm{\phi};\hat{\mathbf{e}}_{i})}k(\bm{\phi};\hat{\mathbf{e}}_{i}|\bm{\phi}';\mathbf{m}')p_{(\bm{\phi}';\mathbf{m}')}(t)\nonumber \\
 & \quad-\gamma_{\bm{\phi}i}'p_{(\bm{\phi};\hat{\mathbf{e}}_{i})}(t)+\gamma_{\bm{\phi}i}'p_{(\bm{\phi};\hat{\mathbf{e}}_{0})}(t),\label{eq:vib-purcell}\\
\frac{dp_{(\bm{\phi};\hat{\mathbf{e}}_{0})}(t)}{dt} & =-k_{(\bm{\phi};\hat{\mathbf{e}}_{0})}^{(\text{out})}p_{(\bm{\phi};\hat{\mathbf{e}}_{0})}(t)\nonumber \\
 & \quad+\sum_{(\bm{\phi}';\mathbf{m}')\neq(\bm{\phi};\hat{\mathbf{e}}_{0})}k(\bm{\phi};\hat{\mathbf{e}}_{0}|\bm{\phi}';\mathbf{m}')p_{(\bm{\phi}';\mathbf{m}')}(t)\nonumber \\
 & \quad-\left(\sum_{j=1}^{2}\gamma_{\bm{\phi}j}'\right)p_{(\bm{\phi};\hat{\mathbf{e}}_{0})}(t)+\sum_{j=1}^{2}\gamma_{\bm{\phi}j}'p_{(\bm{\phi};\hat{\mathbf{e}}_{j})}(t),\label{eq:cav-purcell}
\end{align}
where 
\begin{equation}
\gamma_{\bm{\phi}i}'=\frac{4g^{2}\left(k_{(\bm{\phi};\hat{\mathbf{e}}_{0})}^{(\text{out})}+k_{(\bm{\phi};\hat{\mathbf{e}}_{i})}^{(\text{out})}\right)}{4\Delta^{2}+\left(k_{(\bm{\phi};\hat{\mathbf{e}}_{0})}^{(\text{out})}+k_{(\bm{\phi};\hat{\mathbf{e}}_{i})}^{(\text{out})}\right)^{2}}.\label{eq:gamma'_phi-i}
\end{equation}
Eq. (\ref{eq:gamma'_phi-i}) is in line with standard expressions
for the Purcell factor.\citep{KavokinBook,Hummer2013} Eqs. (\ref{eq:vib-purcell})-(\ref{eq:cav-purcell})
reveal that the dynamics under weak light-matter coupling is identical
to the bare dynamics [Eq. (\ref{eq:master-eqn_k^out})], except that vibrational mode $i$ incoherently
exchanges energy with the cavity at rate $\gamma_{\bm{\phi}i}'$.
Thus, the master equation for the weak coupling regime is that of
the bare case but with the following rates for relaxation between
vibrational and cavity excitations:
\begin{equation}
k(\bm{\phi};\hat{\mathbf{e}}_{i}|\bm{\phi};\hat{\mathbf{e}}_{0})=k(\bm{\phi};\hat{\mathbf{e}}_{0}|\bm{\phi};\hat{\mathbf{e}}_{i})=\gamma_{\bm{\phi}i}',\qquad i=1,2.\label{eq:k_purcell}
\end{equation}

We reiterate that this result rests on the condition that at least
one of cavity and vibrational excitations decays on a timescale faster
than that of the reaction dynamics. In all our simulations involving
weak light-matter coupling, we only consider cases where this separation
of timescales is satisfied, and so we use the kinetic framework described
by Eq. (\ref{eq:k_purcell}) and the immediately preceding text. 

\bibliographystyle{apsrev4-1}
\bibliography{cavityLeakage-gsRxns}

\end{document}